\documentclass{emulateapj}

\usepackage{epsfig}
\usepackage[]{natbib}

\newcommand{\eg}{{\rm e.g., }}
\newcommand{\ie}{{\rm i.e., }}

\newcommand{\Mo}{\ensuremath{M_\odot}}

\newcommand{\asec}{\ensuremath{\arcsec}}

\newcommand{\kmsMpc}{~\ensuremath{{\rm km\ s}^{-1}\ {\rm Mpc}^{-1}}}

\newcommand{\Hnought}{\ensuremath{{\rm H}_0}}

\newcommand{\zprime}{\ensuremath{z^{\prime}}}

\shorttitle{SpARCS : Southern Fields}
\shortauthors{Wilson et al.}

\begin{document}

\title{SPECTROSCOPIC CONFIRMATION OF A MASSIVE RED-SEQUENCE SELECTED GALAXY CLUSTER AT $z=1.34$ IN THE SPARCS-SOUTH CLUSTER SURVEY}

\author{Gillian Wilson\altaffilmark{1}, Adam Muzzin\altaffilmark{2},  H. K. C. Yee\altaffilmark{3}, Mark Lacy\altaffilmark{4}, Jason Surace\altaffilmark{4}, David Gilbank\altaffilmark{5}, Kris Blindert\altaffilmark{6},  Henk Hoekstra\altaffilmark{7,8,9}, Subhabrata Majumdar\altaffilmark{10}, Ricardo Demarco\altaffilmark{1}, Jonathan P. Gardner\altaffilmark{11}, Michael D. Gladders\altaffilmark{12}, and Carol Lonsdale\altaffilmark{13,14}
}

\altaffiltext{1}{Department of Physics and Astronomy, University of California, Riverside, CA 92521; gillianw@ucr.edu}

\altaffiltext{2}{Department of Astronomy, Yale University, New Haven, CT, 06520-8101}

\altaffiltext{3}{Department of Astronomy and Astrophysics, University of Toronto, 50 St. George St., Toronto, Ontario, Canada, M5S 3H4}

\altaffiltext{4}{Spitzer Science Center, California Institute of Technology, 220-6, Pasadena, CA 91125}

\altaffiltext{5}{Astrophysics and Gravitation Group, Department of Physics and Astronomy, University Of Waterloo, Waterloo, Ontario, Canada, N2L 3G1}

\altaffiltext{6}{Max Planck Institute for Astronomy Koenigstuhl 17, 69117, Heidelberg, Germany}

\altaffiltext{7}{Department of Physics and Astronomy, University of Victoria, Victoria, BC V8P 5C2, Canada}

\altaffiltext{8}{Leiden Observatory, Leiden University, PO Box 9513, 2300RA Leiden, The Netherlands}

\altaffiltext{9}{Alfred P. Sloan Fellow}

\altaffiltext{10}{Department of Astronomy and Astrophysics, Tata Institute of Fundamental Research (TIFR), Homi Bhabha Road, Mumbai, India}

\altaffiltext{11}{Goddard Space Flight Center, Code 665, Laboratory for Observational Cosmology, Greenbelt MD 20771}

\altaffiltext{12}{Department of Astronomy and Astrophysics, University of Chicago, 5640 South Ellis Avenue, Chicago, IL 60637}

\altaffiltext{13}{Infrared Processing and Analysis Center, California Institute of Technology, 220-6, Pasadena, CA 91125}

\altaffiltext{14}{North American ALMA Science Center, NRAO Headquarters, 520 Edgemont Road, Charlottesville, VA 22903}

\begin{abstract}

The Spitzer Adaptation of the Red-sequence Cluster Survey (SpARCS) is a \zprime-passband imaging survey, consisting of deep ($\zprime \simeq 24$ AB)
observations made from both hemispheres using
the CFHT 3.6m and CTIO 4m  telescopes. The survey was designed with the primary aim of detecting
galaxy clusters at $z~>~1$. In tandem with pre-existing
 $3.6 \micron$ observations from the 
\emph{Spitzer} Space Telescope SWIRE Legacy Survey, 
SpARCS detects clusters  using an infrared adaptation of the two-filter red-sequence cluster technique. The total effective area of
the SpARCS cluster survey is 41.9 deg$^2$. 
In this paper, we provide an overview of the 13.6 deg$^2$ Southern CTIO/MOSAICII observations. 
The 28.3 deg$^2$ Northern CFHT/MegaCam observations 
are summarized in a companion paper by \citet{muzzin-08b}. In this paper, we also report spectroscopic confirmation of SpARCS J003550-431224, a very rich galaxy cluster at
$z=1.335$, discovered in the ELAIS-S1 field. To date, this is the highest spectroscopically confirmed redshift for a galaxy 
cluster discovered using the red-sequence technique. Based on nine confirmed members, SpARCS J003550-431224 has
a preliminary velocity dispersion of $1050 \pm 230$ km s$^{-1}$.
With its proven capability for efficient cluster detection, SpARCS
is a demonstration that we have entered
an era of large, homogeneously-selected $z>1$ cluster surveys.

\end{abstract}


\section{INTRODUCTION}
\label{sec:intro}

Dating back to the pioneering photographic work of the mid-twentieth century \citep{abell-58, zwicky-68},
galaxy cluster surveys have held a special place in the history of Astronomy. Due to the limitations
of photographic plates, however, early optical surveys struggled to discover clusters at redshifts
higher than $z\simeq0.2$. 

Galaxy cluster surveys were revolutionized by the launch of 
a series of powerful X-ray observatories in the 1980's and 1990's.
Firstly \emph{Einstein}, then \emph{ROSAT}, and later the XMM and \emph{Chandra} telescopes proved their capability to detect
clusters out to  
$z=1$.
For example, the wide but relatively shallow 734 deg$^{2}$ \emph{Einstein} Medium-Sensitivity Survey \citep[EMSS]{gioia-90} both provided the base
catalog for the Canadian Network for Observation Cosmology (CNOC1) survey of clusters at $z\sim0.4$ \citep{yec-96}, and discovered
MS 1054-03 at $z=0.83$  \citep{gl-94}, a massive cluster which provided
the first evidence that $\Omega_m$ was significantly $<1$ \citep{donahue-98}. 
The deeper but smaller 48 deg$^{2}$ \emph{ROSAT} Deep Cluster Survey \citep[RDCS]{rosati-98} detected
RX J0848.9+4452 at $z = 1.26$, which was, at the time, one of the most distant clusters to be discovered \citep{rosati-99}.

The advent of large-format charge coupled devices (CCDs) brought renewed interest in carrying out optical cluster surveys \eg the 6 deg$^{2}$  Palomar Distant Cluster Survey \citep[PDCS]{postman-96}, the 16 deg$^{2}$ KPNO survey of \citet{postman-98},
and the 135 deg$^{2}$ Las Campanas Distant Cluster Survey \citep[LCDCS]{gonzalez-01} from which the ESO Distant Cluster 
Survey \citep[EDisCS]{white-05} was selected. All three of these surveys used variants of the matched-filter method to detect clusters.

In an effort to combat ``false positive'' detections caused by line-of-sight projections of unrelated systems
in single-passband optical cluster searches, \citet{gy-00} proposed a two-filter technique.
Their Cluster Red-Sequence (CRS) technique was motivated by the observational fact that galaxy clusters contain a population of
early-type galaxies which follow a tight color-magnitude relation. 
This relation has been shown to have an extremely small scatter (\eg \citealt{bower-92}), even to redshifts as high as $z\sim1.4$ \citep{blakeslee-03, holden-04, lidman-08}.  If two filters which bracket the $4000\AA$ break are used to construct color-magnitude diagrams, early types are always the brightest, reddest galaxies at any redshift, and provide significant contrast from the field.  
The CRS method has been used for the $\sim100$ deg$^{2}$ Red-sequence Cluster Survey (RCS-1, Gladders \& Yee 2005) and is also being used for the next generation $\sim 1000$ deg$^{2}$ RCS-2 survey \citep{yee-07}. Both of these surveys use an $R-\zprime$ filter combination. A variant of the red sequence method (the BCGmax algorithm)
has also been used to detect clusters in the Sloan Digital Sky Survey \citep{koester-07a, koester-07b}. In addition to detecting clusters with a very low false-positive rate, a second important advantage of the CRS 
technique is that it also provides a good photometric estimate of the cluster redshift, accurate to $\Delta z \sim 0.05$ at $z<1$ \citep{gilbank-07}.

Applying the CRS technique to searching for clusters at higher redshift was an obvious next step which
was not feasible until very recently, because of technical limitations.
At $z\sim1.2$, the \zprime\ filter is no longer redward of the rest-frame $4000\AA$ break, requiring
the use of large-format near-infrared cameras which have only recently begun to appear on 4m telescopes \citep{dalton-06,warren-07}.  
Another issue is that the sky itself is bright in the infrared, requiring longer integration times than for optical imaging.

The first real opportunity to systematically detect galaxy clusters at  $z>1$ in large numbers was presented in 2003
with the launch of the InfraRed Array Camera (IRAC; \citealt{fazio-04}) onboard the \emph{Spitzer} Space Telescope \citep{werner-04}. 
Both our own pilot study carried out using the 
3.8 deg$^{2}$ 60s-depth First Look Survey 
(FLS \citealt{lacy-05, wilson-05, muzzin-08a}), 
and that carried out using the 
8.5 deg$^{2}$ 90s-depth  IRAC Shallow Survey
\citep{eisenhardt-04, stanford-05, brodwin-06, eisenhardt-08},  quickly demonstrated the power of IRAC
for $z>1$ cluster detection.
 
Clusters of galaxies are extremely rare and one requires a widefield survey to find the most massive examples.
The largest area \emph{Spitzer} Space Telescope
Survey is the 50 deg$^{2}$ 120s-depth SWIRE Legacy Survey \citep{lonsdale-03}. 
In \S~\ref{sec:obs}, we provide an overview of SpARCS\footnote{http://www.faculty.ucr.edu/$\sim$gillianw/SpARCS}, our \zprime-imaging survey of the six SWIRE fields.
We also summarize the Southern observations.
In \S~\ref{sec:clusdet}, we briefly review our red-sequence detection algorithm and
introduce a rich cluster candidate 
selected in the CTIO ELAIS-S1 field.
We present Gemini/GMOS-S spectroscopic follow-up of SpARCS J003550-431224 
in \S~\ref{sec:spectra}, confirming it to be, at $z=1.34$, the highest redshift cluster yet discovered using the red-sequence technique.
We discuss our main results in  \S~\ref{sec:disc} and conclude in \S~\ref{sec:conc}.
We assume an  $\Omega_{{\rm m}0} = 0.3, \Omega_{\lambda 0} = 0.7$ 
cosmology with $\Hnought = 70$ $\kmsMpc$ throughout.

\section{SpARCS SURVEY IMAGING}
\label{sec:obs}

\subsection{Choice of Passbands}

The Spitzer SWIRE Legacy Survey is a seven passband imaging survey consisting of 
IRAC 3.6, 4.5, 5.8, 8.0 $\micron$  
and Multiband Imaging Photometer for Spitzer (MIPS;  \citealt{rieke-04}) 24, 70, 160 $\micron$ observations.
Full details of the survey design, data processing, ancillary datasets and source catalogs
may be found in \cite{surace-05}. 
For cluster detection SpARCS utilizes
$3.6\micron$, the most sensitive \emph{Spitzer} channel, as its ``red'' filter.

SpARCS utilizes \zprime\ as its  ``blue'' filter. Simulations (\eg\ see \citealt{wilson-06}) showed  a limiting magnitude of $z\sim24$ AB
was required to match the 3.6$\micron$ depth.
It was necessary to obtain our own widefield \zprime-imaging for only five of the six SWIRE fields (Table~\ref{tab:fields}), because observations of the
XMM-LSS field were available from the Canada-France-Hawaii Telescope 
(CFHT)  Legacy Survey\footnote{http://www.cfht.hawaii.edu/Science/CFHTLS/}. Those observations were made either using the  MOSAICII instrument on the 4m Blanco telescope at the Cerro Tololo Inter-Amercian Observatory (CTIO) in the case of the Southern fields (ELAIS-S1 and Chandra-S) or
using  MegaCam 
at the 3.6m CFHT in the case of the  Northern fields (ELAIS-N1, ELAIS-N2 and Lockman).
A summary of the latter observations, and spectroscopic confirmation of two clusters at $z=1.18$ and $z=1.20$ in the ELAIS-N2 field may be found in the companion paper by \cite{muzzin-08b}. 

\subsection{The CTIO Dataset}

\begin{figure}
\plotone{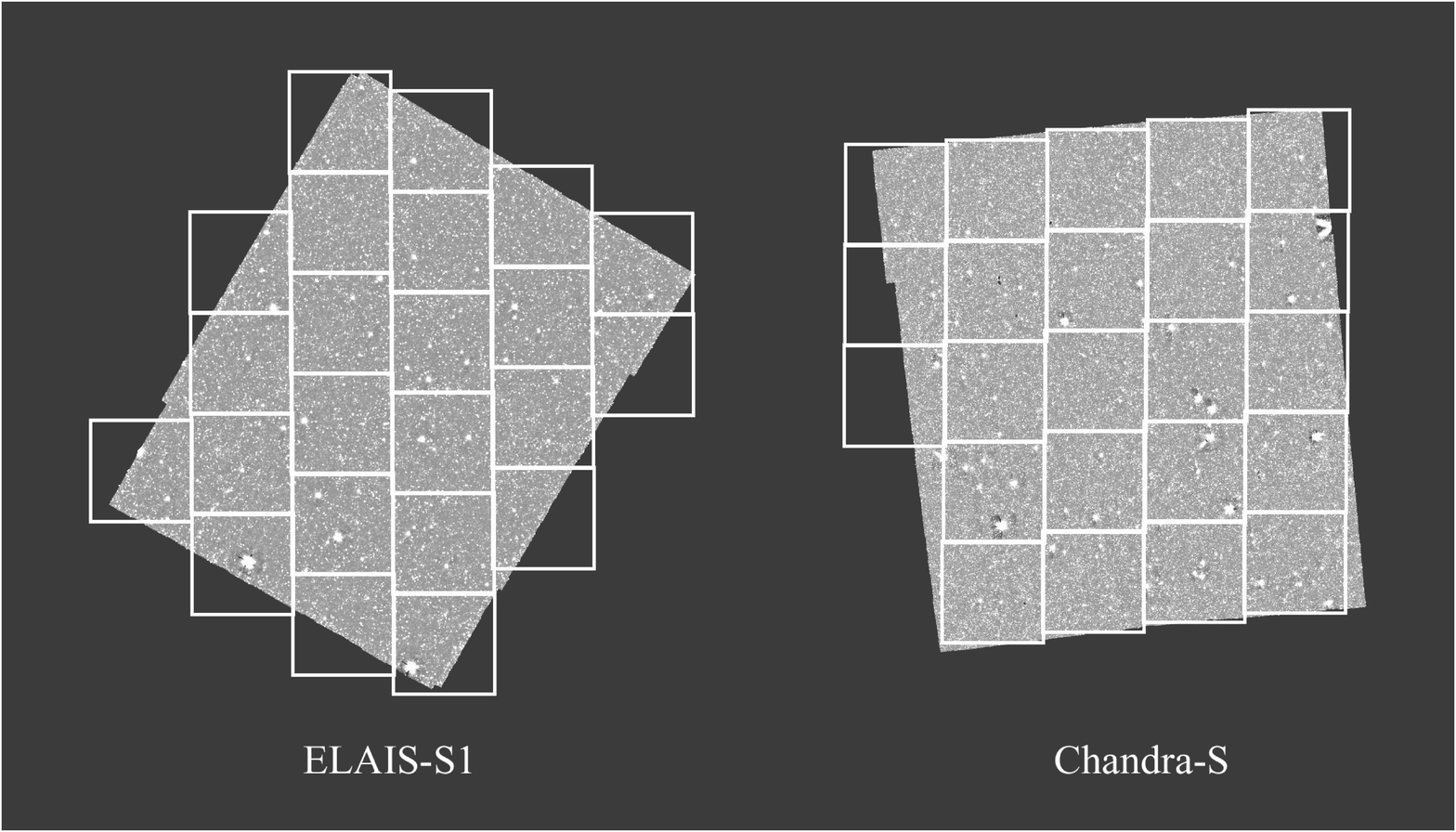}
\caption{
The SWIRE $3.6\micron$ IRAC mosaics are shown in grayscale. The roll angle of the Spitzer Space Telescope on the date of observation
determines the orientation of the IRAC observations. The white squares overlaid show the 
46 CTIO MOSAICII pointings. There are 23 MOSAICII pointings each in the SWIRE 7.1 deg$^{2}$ ELAIS-S1  and 
8.1 deg$^{2}$ Chandra-S fields.
\label{fig:overlay}
}
\end{figure}

The SWIRE IRAC $3.6\micron$ Southern fields are shown in Figure~\ref{fig:overlay}.
The $3.6\micron$ mosaic of the ELAIS-S1 field totals 7.1 deg$^{2}$ and the 
Chandra-S field totals 8.1 deg$^{2}$ (see Table~\ref{tab:fields})

The $8192\times8192$ pixel MOSAIC II camera on the 4m Blanco telescope has a
pixel scale of $0. \arcsec 267$ pixel$^{-1}$, leading to a $36\arcmin \times36 \arcmin$ footprint per pointing. The white squares overlaid on Figure~\ref{fig:overlay} show the 
46 CTIO MOSAICII pointings required to image the ELAIS-S1  and Chandra-S fields.
These pointings were designed to maximize the overlap with the $3.6\micron$  data, but to minimize the overall 
number of pointings by omitting regions with little overlap with the IRAC data. The total area of the \zprime\ observations per field is shown in Table~\ref{tab:fields}.

CTIO observations were made of the ELAIS-S1 and Chandra-S fields using the  \zprime\ filter on a total of 17 nights.
The depth of the \zprime\ data varies from pointing to pointing depending on the seeing and the sky background; however, the mean depth is 24.0 AB (23.5 Vega; $5 \sigma$).
Table~\ref{tab:fields} shows 
the total effective area per field, \ie the total usable area of overlap between the \zprime\ and $3.6 \micron$ datasets once areas of bright star contamination and chips gaps have been excluded. The total effective area of the CTIO fields is 13.6 deg$^{2}$ (6.5 deg$^{2}$ in the ELAIS-S1 and 7.1 deg$^{2}$ in the Chandra-S fields), and the total effective area of the six fields is 41.9 deg$^{2}$.


\section{CLUSTER DETECTION}
\label{sec:clusdet}

We defer a full description of the SpARCS data reduction, cluster candidate detection algorithm and catalogs
to Muzzin et al.\ 2009, in prep.
However, we note that the algorithm used here is almost identical to that 
described in \citet{muzzin-08a}, as applied to the Spitzer First Look 
Survey (FLS; \citealt{lacy-05, wilson-05}). The one important difference is that
\citet{muzzin-08a} used an $R-[3.6]$ color to detect clusters at $0 < z < 1.3$ in the FLS.
The slightly deeper SWIRE exposure time, combined with the $\zprime-[3.6]$ color choice, allows 
SpARCS to detect clusters to higher redshift than possible with the FLS dataset.

\subsection{SpARCS J003550-431224}
\label{ssec:cluster}

From analysis of the ELAIS-S1 field, SpARCS J003550-431224,
shown in Figure~\ref{fig:images}, was identified 
as a high probability rich cluster candidate.
SpARCS J003550-431224 (R.A.: 00:35:49.7, Dec.:-43:12:24.16) has a B$_{\rm gc,R}$\footnote{We use the $\zprime - [3.6]$ vs. $3.6\micron$ red-sequence to determine  B$_{\rm gc,R}$} richness of $1055 \pm 276$ Mpc$^{1.8}$ (for a discussion of B$_{\rm gc}$ and B$_{\rm gc,R}$ see \citealt{yeelc-99} and  \citealt{gy-05}). Based on the empirical calibration of B$_{\rm gc}$  vs. M$_{200}$ determined by \citet{muzzin-07b} in the K-band for the CNOC1 clusters at $z \sim 0.3$, this implies M$_{200}= 5.7 \times 10^{14} \Mo$.

\section{SPECTROSCOPY}
\label{sec:spectra}

Spectroscopic follow-up of  SpARCS J003550-431224 was obtained 
with the Gemini Multi-Object Spectrograph on the Gemini South telescope (GMOS-S) 
in queue mode (program ID GN-2007B-Q16). We used $1.0 \asec$ wide slits.
We used GMOS with the R150 grism, blazed at 7170$\AA$.  This provided a spectral resolving power of R = 631 which corresponds to a resolution of $11\AA$, or 280 km s$^{-1}$ at the redshift of the cluster. 

We observed a single mask for 10 hrs. The mask was observed using the OG515 filter 
which blocks light blueward of $5150\AA$.  The central wavelength of the grating was moved between $7380\AA$, $7500\AA$, and $7620\AA$ to ``dither'' in the dispersion direction and fill in GMOS chip gaps.
We used nod-$\&$-shuffle in band-shuffle mode. This is slightly less efficient than micro-shuffle mode but allows one to maximize the number of slits 
in a small area, such as is the case for a distant cluster with a high surface density of red-sequence galaxies concentrated in the cluster core
(see Figure~\ref{fig:images}). There
were 26 slits on the mask, including three alignment stars.
Slits were placed on galaxies with priorities from 1 to 5. Priority 1 (single digit ID $\#$'s in Table~\ref{tab:spectra}) was galaxies with colors within 0.6 magnitude of the red sequence, and with $[3.6] < 17.0$ (Vega). Priority 2 (ID $\#$'s in the 1000's) was galaxies with colors within 0.6 magnitude of the red sequence, and
with $17.0 < [3.6] < 18.0$. Priority 3 (ID $\#$'s in the 2000's) was galaxies with colors bluer than the red sequence  by $0.6-1.0$  magnitude, and with $[3.6] < 17.5$. Priority 4 (ID $\#$'s in the 3000's) was galaxies with colors bluer than the red-sequence by $1.0-1.5$ magnitude, and with $[3.6] < 17.5$. Priority 5 (ID $\#$'s in the 4000's) was the lowest priority and included all galaxies with magnitude $16.9 < [3.6] < 18.0$. Priorities 1 to 4 roughly correspond to bright red-sequence, faint red-sequence, blue cloud, and extreme blue cloud galaxies respectively. Each exposure was 30 minutes in duration.
The 20 exposures used nod cycles of 60s integration time per cycle and were offset by a few arcseconds using the on-chip dithering option.

\subsection{Data Reduction}

The data were reduced using standard GeminiIRAF routines to bias-subtract the data. 
The iGDDS package \citep{abraham-04} was used to interactively trace the 2D spectra and extract 1D spectra.
Wavelength-calibration for each extracted spectrum was performed-using bright sky lines from the unsubtracted image, also
with the iGDDS software. Wavelength solutions typically have an r.m.s.\ $ < 0.5\AA$. We determined a relative flux calibration curve using a long slit 
observation of the standard star EG21.
We determined redshifts by identifying prominent spectra features such as Calcium H+K lines at 3934$\AA$ and 3968 $\AA$ and the [OII] $\lambda \lambda 3727$ doublet  using the iGDDS code. A few of the spectra also show some Balmer series lines.

\begin{figure}
\plotone{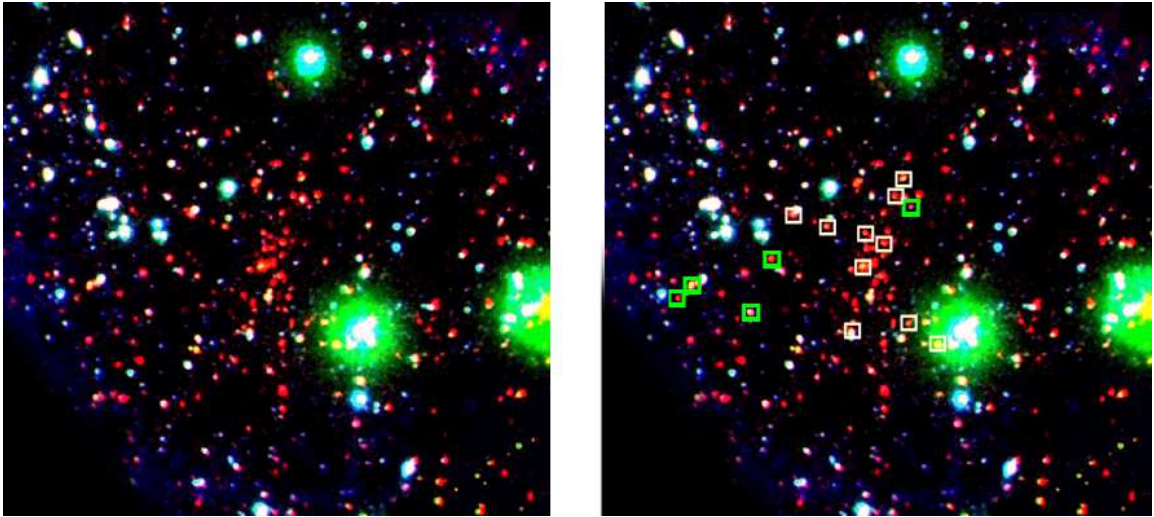}
\caption{
$r^{\prime}\zprime[3.6]$ color composites of J003550-431224 are shown in both panels.  The FOV is 1.5 Mpc at the cluster redshift.
The white (green) boxes overlaid on the right panel
show the 10 cluster members (5 foreground/background galaxies) with spectroscopically-confirmed redshifts from Gemini/GMOS-S (see Table~\ref{tab:spectra}).
\label{fig:images}
}
\end{figure}

\begin{figure}
\plotone{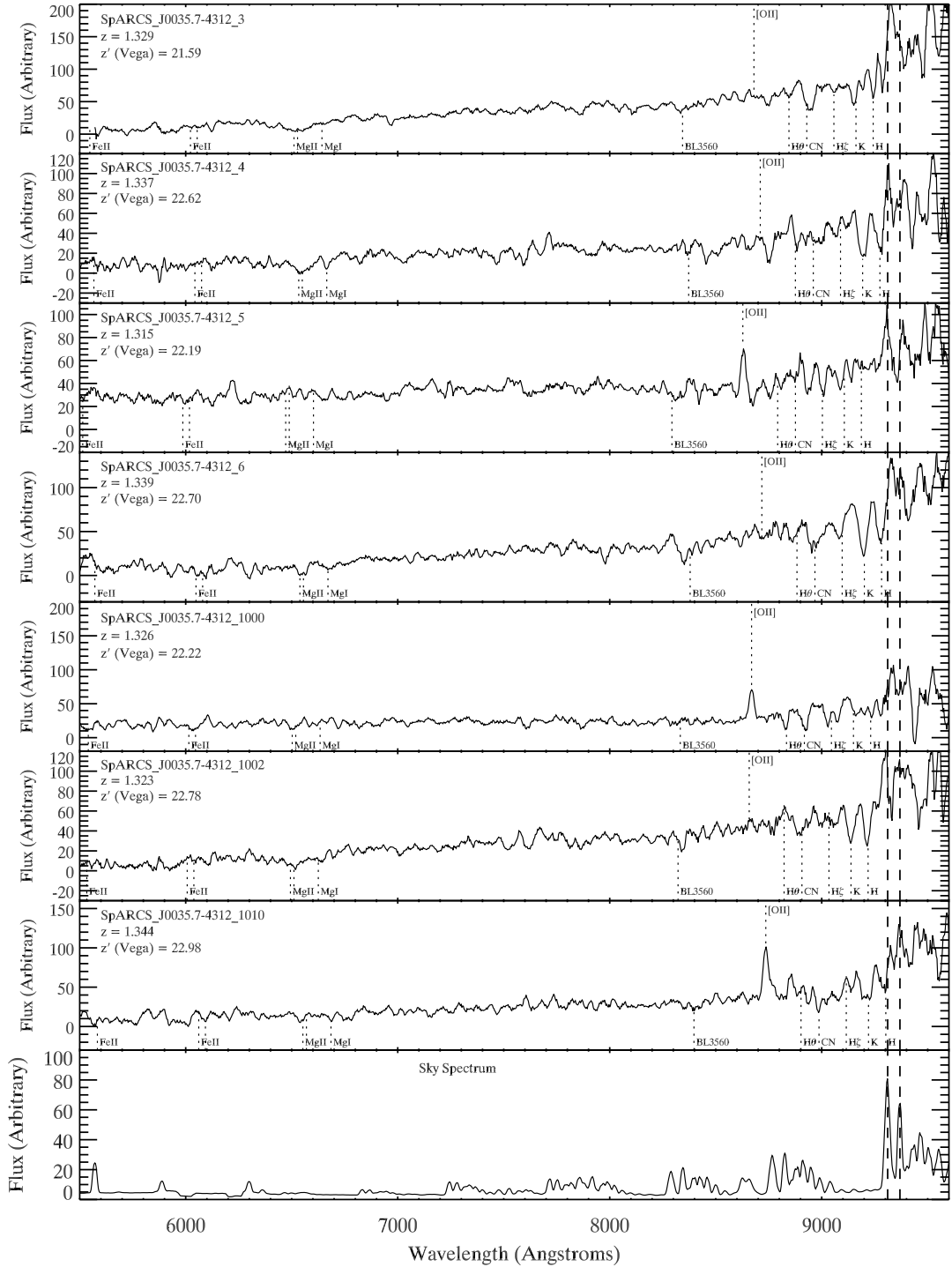}
\caption{
Spectra for a subsample of seven galaxies in cluster SpARCS J003550-431224 (see Table~\ref{tab:spectra}). The spectra have been smoothed with a 7-pixel ($11 \AA$) boxcar. The identified spectral features
are marked. The lowermost panel shows the typical sky spectrum.
\label{fig:spectra}
}
\end{figure}

Spectra were obtained for 15 of the 23 photometrically selected galaxies with quality sufficient
for determining redshifts: the other eight were deemed too faint for reliable identification of spectral 
features. Table~\ref{tab:spectra} shows ten galaxies, which were deemed likely to 
be cluster members based upon the value of their spectroscopic redshifts (although see \S~\ref{sec:disc}).
These galaxies are indicated by white boxes in the right panel of Figure~\ref{fig:images}.
Some examples of the spectra of confirmed members are shown in Figure~\ref{fig:spectra}.
These spectra have been smoothed by a 7-pixel boxcar (which produces a resolution equal to that of the spectrograph).


A total of five galaxies were determined 
to be foreground or background sources. These galaxies are also listed in Table~\ref{tab:spectra} and indicated by green boxes in the right panel of Figure~\ref{fig:images}.
The histogram in the left panel of Figure~\ref{fig:hist} shows the spectroscopic redshifts of all 15 galaxies.
The histogram in the right panel 
shows only the redshifts of the 10 likely cluster members. 

\section{DISCUSSION}
\label{sec:disc}

\begin{figure}
\plotone{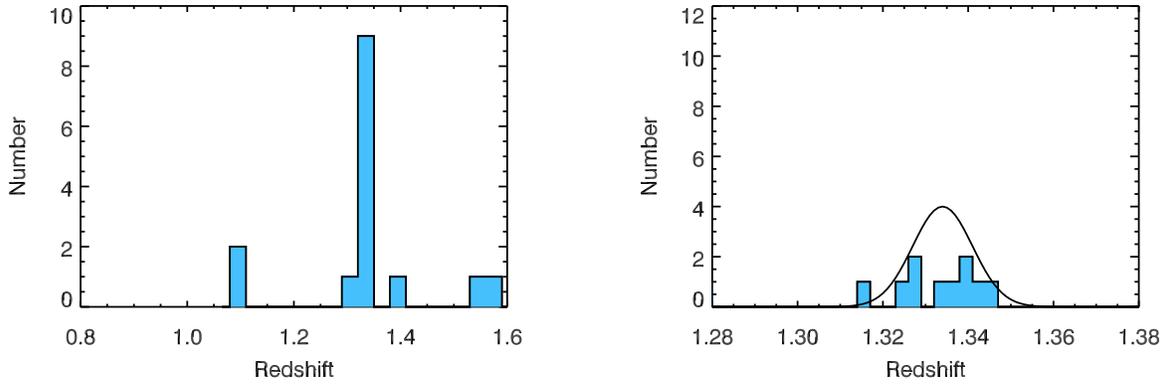}
\caption{
Histogram showing 15 galaxies with confirmed redshifts (left).  Ten galaxies are confirmed as cluster members (right) and five are confirmed as foreground or background sources.  See Table~\ref{tab:spectra} for further details. A Gaussian with an r.m.s. of 1050 km s$^{-1}$  (see \S~\ref{sec:disc}) has been overlaid
\label{fig:hist}
}
\end{figure}

\begin{figure}
\plotone{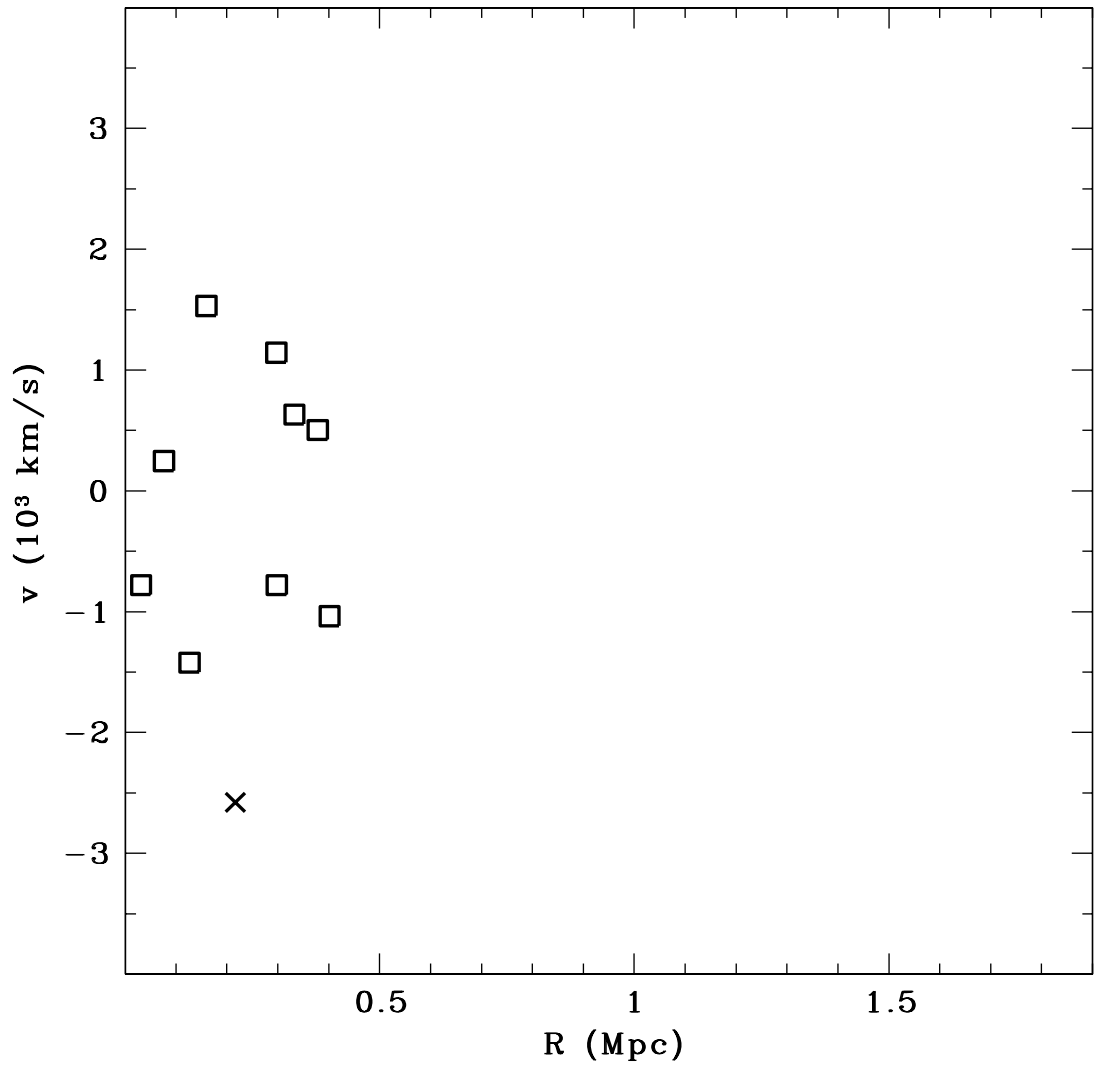}
\caption{
Galaxy velocities relative to the mean velocity, as a function of radius. The galaxy marked with an ``x'' (ID \# 5 at $z=1.315$ in Table~\ref{tab:spectra}) is more likely to a near-field object than a member of the cluster and was not used in the computation of the mean redshift or velocity dispersion.
\label{fig:vd}
}
\end{figure}

Before calculating a redshift and velocity dispersion for SpARCS J003550-431224,
we first checked for near-field interlopers using the code of \citet{blindert-06}.
This employs a modified version of the \citet{fadda-96} shifting-gap technique, and uses
both galaxy position and velocity information.
Figure~\ref{fig:vd} shows galaxy velocities relative to the mean velocity, as a function of radius. The galaxy marked with an ``x'' (ID 5 at $z=1.315$ in Table~\ref{tab:spectra}) was 
identified as being more likely to be a near-field object than a member of the cluster and was not used in the computation of the velocity dispersion or the mean redshift of the cluster.
(This is also the lowest redshift galaxy in the right panel of Figure~\ref{fig:hist}).
A mean redshift of $z=1.335$ and a velocity dispersion of $1050 \pm 230$ km s$^{-1}$ was
 calculated for SpARCS J003550-431224. The uncertainty on the latter was determined using Jackknife 
resampling of the data.

It is, as yet, unclear as to whether SpARCS J003550-431224 is a relaxed cluster or a system in the process of collapsing.
The distribution in Figure~\ref{fig:hist} (right panel) appears non-Gaussian and there is some evidence of bimodality.
For comparison, a Gaussian with an r.m.s. of 1050 km s$^{-1}$ has been overlaid on the right panel of  Figure~\ref{fig:hist}. 
The  velocity dispersion of $1050 \pm 230$ km s$^{-1}$ should certainly be considered ``preliminary'', based, as it is, upon
only nine members.
However, if one does adopt this value of velocity dispersion, one can use it to calculate a dynamical estimate
of r$_{200}$ (the radius at which the
mean interior density is 200 times the critical density, $\rho_{c}$) using the equation
of \citet{carlberg-97}:

\begin{equation}
r_{200} = \frac{\sqrt{3}\sigma}{10H(z)},
\end{equation}
where H($z$) is the Hubble parameter at the redshift of the cluster.
This gives a  value of  r$_{200}$ = 1.2 $\pm$ 0.3 Mpc. From this, 
the dynamical mass M$_{200}$, the mass contained within r$_{200}$, can also be inferred
using:
 
\begin{equation}
M_{200} = \frac{4}{3}\pi r_{200}^3 \cdot 200\rho_{c},    
\end{equation}

We estimate a dynamical mass  of M$_{200}= (9.4 \pm 6.2) \times 10^{14} \Mo$ for J003550-431224, which is 
in good agreement with, albeit slightly larger than, the value of  M$_{200}= 5.7 \times 10^{14} \Mo$, which was 
calculated based upon its richness. Further spectroscopy is certainly warranted, but the preliminary evidence
suggests that SpARCS J003550-431224 is a massive cluster, perhaps the most massive cluster at $z>1$ in
our 42 deg$^{2}$ survey.

\begin{figure}
\epsscale{1.0}
\plotone{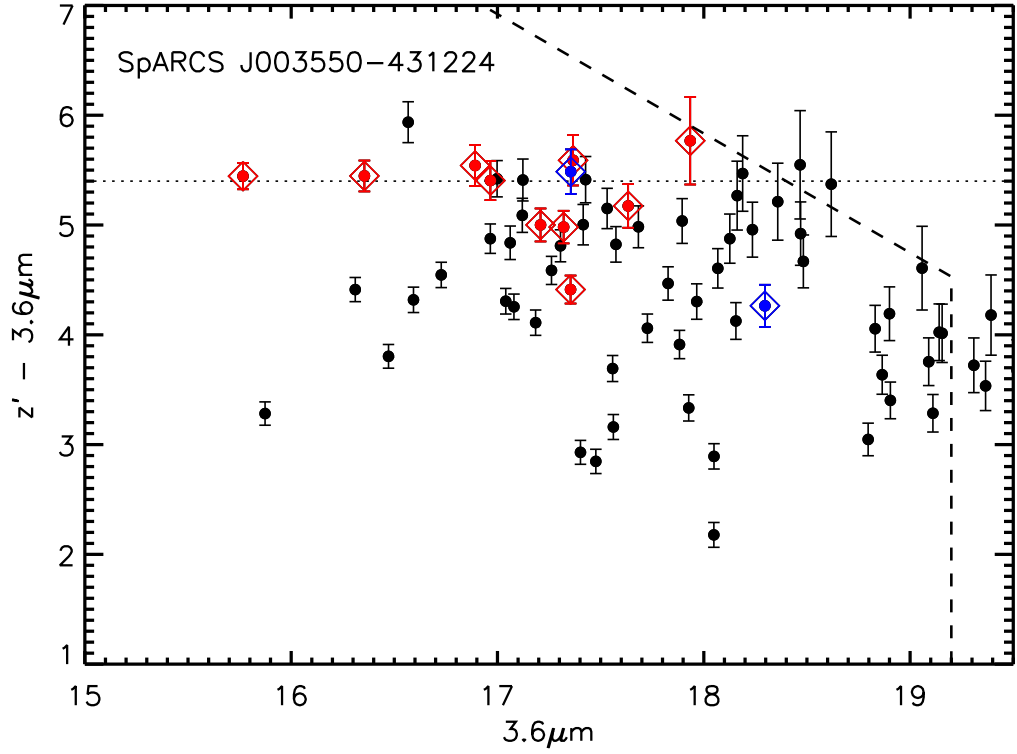}
\caption{
\zprime-[3.6] versus [3.6] color-magnitude diagram for SpARCS J003550-431224.
The black circles are all the galaxies contained within a circle of radius 550 kpc ($65\asec$)
at the cluster redshift.
The red diamonds show the ten spectroscopically confirmed cluster members. The blue
diamonds show the two (of the five) confirmed foreground/background galaxies which fall within the 550 kpc radius. 
The dotted line indicates the nominal red-sequence color for this cluster ($\zprime-[3.6]=5.4$).
See \S~\ref{sec:disc} for  a discussion.
\label{fig:colmag}
}
\end{figure}

The color-magnitude diagram for all galaxies in a 550 kpc ($65\asec$) radius of the cluster center
is shown in Figure~\ref{fig:colmag}. 
The red diamonds show the 10 cluster members from Table~\ref{tab:spectra}. The blue
diamonds show the two (of five) confirmed foreground/background galaxies that fall within the 550 kpc radius. 

Based on its red-sequence color of $\zprime-[3.6]=5.4$ (shown as dotted line in Figure~\ref{fig:colmag}),
SpARCS J003550-431224 was originally assigned a photometric redshift of $z=1.57$.
This was based on \citet[BC03]{bc-03} models assuming a formation redshift $z_{f}=4$.
If one assumes a formation redshift of $z_{f}=10$, then the $\zprime-[3.6]=5.4$ color would cause one to infer a photometric redshift
of $z=1.39$ for SpARCS J003550-431224, which is reasonably consistent with the spectroscopic value. 
One possibility  for the overestimate of the photometric redshift is that the BC03 models are too blue at young ages, as suggested by  \citet{maraston-05}.
Spectroscopy of additional $z\sim1.35$ SpARCS clusters 
will determine whether SpARCS J003550-431224's red-sequence color is representative for this redshift, or if there
is something atypical about this particular cluster.

\section{CONCLUSIONS}
\label{sec:conc}

In this paper, we reported the discovery of SpARCS J003550-431224 in the ELAIS-S1 field, a very rich galaxy cluster at $z=1.335$, with a preliminary velocity dispersion of $1050 \pm 230$ km s$^{-1}$. Several other rich $z>1$ clusters in the SpARCS survey have also been spectroscopically confirmed
(\citealt{muzzin-08b}, Demarco et al.\ 2008, in prep), with no false positives.


The FLS, IRAC Shallow Cluster Survey (ISCS) and SpARCS have demonstrated the power and potential of widefield infrared observations from space for $z>1$ cluster surveys. Both the FLS and SpARCS surveys are very similar to the RCS-1 and RSC-2 surveys, except that they utilize an optical-infrared adaptation of the CRS technique (the ISCS uses a multipassband photometric technique).
SpARCS, in particular, is a demonstration that $z>1$ cluster detection is now routine. With its two filter red-sequence detection algorithm,  
it is both highly efficient and robust against line-of-sight projections.
At  42 deg$^2$, the SpARCS cluster survey is also currently 
the only $z>1$ survey sufficiently large to select  clusters \emph{both} as a function of redshift and richness.

It is noteworthy that all the $z>1$ SpARCS clusters were discovered in only 120s per pointing of IRAC observations.
Since clusters are rare, widefield cluster searches are required to discover the most massive examples.
To discover a representative sample of massive structures at $z>1.5$ would require an even wider survey, of several hundred square
degrees, a proposition which would be feasible during the lifetime of the \emph{Spitzer} Space Telescope Warm Mission  \citep{gardner-07, stauffer-07}.

\acknowledgements 

The authors wish to thank the staff of The Cerro Tololo Inter-American Observatory for their invaluable assistance, 
without which this work would not have been possible.  CTIO is operated by the Association of Universities for 
Research in Astronomy, under contract with the National Science Foundation.

Based on observations obtained at the Gemini Observatory, which is operated by the
Association of Universities for Research in Astronomy, Inc., under a cooperative agreement
with the NSF on behalf of the Gemini partnership: the National Science Foundation (United
States), the Science and Technology Facilities Council (United Kingdom), the
National Research Council (Canada), CONICYT (Chile), the Australian Research Council
(Australia), MinistŽrio da Cincia e Tecnologia (Brazil) and SECYT (Argentina)

This work is based in part on archival data obtained with the Spitzer Space Telescope, which is operated by the Jet Propulsion Laboratory, California Institute of Technology under a contract with NASA. Support for this work was provided by an award issued by JPL/Caltech.
GW acknowledges support from the College of Natural and Agricultural Sciences at UCR.

\newpage


\clearpage

\clearpage

\clearpage

\clearpage

\clearpage

\clearpage

\clearpage

\begin{deluxetable}{rccrrr}
\tablewidth{0pt}
\tablecaption{The SpARCS Fields. \label{tab:fields}}
\tablewidth{0pt}
\tablehead{\colhead{Field} & \colhead{R.A.} & \colhead{Decl.} & \colhead{SWIRE $3.6\micron$} & \colhead{SpARCS $\zprime$} & \colhead{Usable} \
\\
 \colhead{} & \colhead{J2000 (Deg.)} & \colhead{J2000 (Deg.)} & \colhead{Area(deg$^{2}$)} & \colhead{Area(deg$^{2}$)} & \colhead{Area(deg$^{2}$)} \
}
\startdata
\\
ELAIS-S1                       & 00:38:30 & -44:00:00 &  7.1  & 8.3  & 6.5  \
\\
XMM-LSS                        & 02:21:20 & -04:30:00 &  9.4  & 11.7 & 7.3 \
\\
Chandra-S                       & 03:32:00 & -28:16:00 &  8.1  & 7.9 & 7.1  \
\\
Lockman                        & 10:45:00 & +58:00:00 & 11.6  & 12.9 & 9.7 \
\\
ELAIS-N1                       & 16:11:00 & +55:00:00 &  9.8  & 10.3 & 7.9 \
\\
ELAIS-N2                       & 16:36:48 & +41:01:45 &  4.4  &  4.3 & 3.4 \
\\
\hline
\\
Total                          &          &          &  50.4  & 55.4 & 41.9 \
\\
\enddata
\end{deluxetable}

\clearpage

\begin{deluxetable}{lcccc}
\tabletypesize{\footnotesize}
\scriptsize
\tablecolumns{5}
\tablecaption{Spectroscopic Redshifts in the Field of SpARCSJ0035.7-4312 \label{tab:spectra}}
\tablewidth{2.0in}
\tablehead{\colhead{ID} & \colhead{R.A.} & \colhead{Decl.} & \colhead{z$^{\prime}$} & \colhead{$z$} \
\\
 \colhead{} & \colhead{J2000 (Deg.)} & \colhead{J2000 (Deg.)} &\colhead{Mag Vega} & \colhead{} \
}
\startdata
\multicolumn{5}{c}{Members}\
\\
\hline
 3      &     0.597095   &    -43.20612    &     21.59  &   1.329 \
\\   
 4      &     0.596806   &    -43.20263    &     22.62  &   1.337 \
\\  
 5      &     0.598068   &    -43.19822    &     22.19  &   1.315 \
\\
 6      &     0.596565   &    -43.19285    &     22.70  &   1.339 \
\\
 1001   &     0.596076   &    -43.21811    &     22.22  &   1.327 \
\\   
 1002   &     0.596473   &    -43.21479    &     23.69  &   1.329 \
\\
 1009   &     0.597075   &    -43.20102    &     22.78  &   1.324 \
\\   
 1010   &     0.597606   &    -43.20000    &     22.28  &   1.347 \
\\ 
 1012   &     0.596664   &    -43.19551    &     22.98  &   1.344 \
\\
 3002   &     0.597252   &    -43.21589    &     21.81  &   1.340 \
\\ 
\hline
  \multicolumn{5}{c}{Foreground/Background} \
\\
\hline
 1      &     0.599490   &    -43.20939    &    22.33  &   1.100 \
\\ 
 1007   &     0.598371   &    -43.20495    &     22.84  &   1.394 \
\\ 
 2010   &     0.599674   &    -43.21098    &     22.98  &   1.104 \
\\ 
 3003   &     0.598673   &    -43.21307    &     21.62  &   1.550 \
\\   
 4001   &     0.596434    &    -43.19699    &     22.68  &   1.581 \
\\  
\enddata
\end{deluxetable}

\end{document}